\documentclass[runningheads]{llncs}
\usepackage{graphicx}
\usepackage{xcolor}
\usepackage{algorithm}
\usepackage[noend]{algorithmic}
\usepackage{bbm}
\usepackage{bm}
\usepackage{booktabs}
\usepackage[symbol]{footmisc}
\usepackage{adjustbox}
\usepackage{caption}
\usepackage{hyperref}
\usepackage[misc]{ifsym}
\usepackage{ulem}

\begin{document}
%
\title{DiffCMR: Fast Cardiac MRI Reconstruction with Diffusion Probabilistic Models}
\titlerunning{DiffCMR}
%
%
\author{Tianqi Xiang\inst{1\footnotemark[2]} \and Wenjun Yue\inst{1,2\footnotemark[2]} 
\and Yiqun Lin\inst{1} \and Jiewen Yang\inst{1} \and \\
Zhenkun Wang \inst{3} \and Xiaomeng Li\inst{1\footnotemark[1]}}
\authorrunning{T. Xiang et al.}
\institute{The Hong Kong University of Science and Technology, Hong Kong, China \\ \email{eexmli@ust.hk}\\ \and HKUST Shenzhen Research Institute, Shenzhen, China \and Southern University of Science and Technology, Shenzhen, China}
%
\maketitle

\renewcommand{\thefootnote}{\fnsymbol{footnote}}
\footnotetext[2]{These authors contributed equally.}
\footnotetext[1]{Corresponding author.}

\newcommand{\yq}[1]{{\color{red}{#1}}}
\newcommand{\wj}[1]{{\color{blue}{#1}}}
\newcommand{\Q}[1]{{\color{magenta}{#1}}}

\newcommand{\xmli}[1]{{\color{red}{XM:#1}}}

\begin{abstract}

Performing magnetic resonance imaging (MRI) reconstruction from under-sampled k-space data can accelerate the procedure to acquire MRI scans and reduce patients' discomfort. The reconstruction problem is usually formulated as a denoising task that removes the noise in under-sampled MRI image slices. 
Although previous GAN-based methods have achieved good performance in image denoising, they are difficult to train and require careful tuning of hyperparameters. In this paper, we propose a novel MRI denoising framework DiffCMR by leveraging conditional denoising diffusion probabilistic models. Specifically, DiffCMR perceives conditioning signals from the under-sampled MRI image slice and generates its corresponding fully-sampled MRI image slice. 
During inference, we adopt a multi-round ensembling strategy to stabilize the performance. We validate DiffCMR with cine reconstruction and T1/T2 mapping tasks on MICCAI 2023 Cardiac MRI Reconstruction Challenge (CMRxRecon) dataset. Results show that our method achieves state-of-the-art performance, exceeding previous methods by a significant margin.
Code is available at \url{https://github.com/xmed-lab/DiffCMR}.

\keywords{Under-sampled MRI \and Cardiac MRI \and MRI reconstruction \and Denoising diffusion probabilistic models}
\end{abstract}
\section{Introduction}

Magnetic resonance imaging (MRI) is an important non-invasive imaging technique that visualizes internal anatomical structures without radiation doses. However, the acquisition time for MRI is significantly longer than X-ray-based imaging since a series of data points should be collected in the k-space. 
Particularly, cardiac magnetic resonance imaging (CMR) requires more time for acquisition as the heart beats uncontrollably and the data acquisition period should cover several heartbeat cycles.
Long scanning time for MRI usually brings discomfort and stress to patients, which may induce artifacts in the MRI reconstruction process. In this work, we study the under-sampled MRI reconstruction that sparsely samples k-space data for image reconstruction, which is one of the ways to accelerate MRI acquisition.

The reconstruction from under-sampled k-space data is an ill-posed problem, which is challenging and has received much attention. In the past, compressive sensing~\cite{CS} is used to solve an optimization problem that seeks to find the most compressible representation that is consistent with the under-sampled k-space data.
With the development of deep learning, the reconstruction problem is usually formulated as a denoising task that removes the artifacts in under-sampled MRI images. For example, encoder-decoder-based methods~\cite{end2end,end2end_2,end2end_3} are proposed to learn a mapping from under-sampled images to fully-sampled images. GAN-based methods~\cite{cgan} introduce a discriminator network that learns to identify the differences between the generated and real images, which can further improve image quality. However, GAN-based methods are difficult to train and require careful tuning of hyperparameters.

Recently, denoising diffusion probabilistic models~\cite{ddpm,ddim} (DDPMs) are proposed to use a series of transformations to increase the complexity of the generated output iteratively. Compared with GAN-based methods, DDPMs have been shown to be more stable during training and demonstrate outstanding performance on a variety of computer vision tasks~\cite{beat}, including image synthesis~\cite{ddpm,ddim,cddpm}, inpainting~\cite{inpaint,repaint}, segmentation~\cite{segdiff,medsegdiff}, and denoising~\cite{xiang2023ddm,restore,ddpmGdenoise}.
To this end, we leverage conditional DDPMs into under-sampled MRI reconstruction. Specifically, we employ a conditional DDPM that generates fully-sampled MRI slices with a conditioning signal from the input under-sampled MRI slices and further adopt multi-round inference ensembling to stabilize the denoising process.

To summarize, the main contributions of this work include 1.) we propose DiffCMR for fast MRI reconstruction from under-sampled k-space data by leveraging conditional diffusion models; 2.) extensive experiments are conducted on MICCAI 2023 CMRxRecon dataset, showing DiffCMR's state-of-the-art performance that outperforms previous methods by a large margin.

\section{Methodology}
\subsection{Problem Definition}

In this work, we formulate the reconstruction of high-quality MRI from under-sampled k-space data as a denoising task. Specifically, inverse Fast Fourier Transform (iFFT) is performed to transform fully-sampled and under-sampled k-space data into 2D image slices, which are referred to as fully-sampled and under-sampled image slices in subsequent sections.
We denote the set of generated image slices as $D^{(k)}= \{(I_{i,k}^u, I_{i,k}^f)_{i=1}^{N_k}\}$, where $k$ refers to three different acceleration factors; $N_k$ is the number of samples generated from the raw data with acceleration factor $k$; $I_{i,k}^u$ indicates the $i$-th under-sampled image slice with acceleration factor $k$ and $I_{i,k}^f$ is the corresponding fully-sampled image slice.
For each acceleration factor $k$, we aim to train a denoising model $\phi_\theta^{(k)}$ with the dataset $D^{(k)}$, which has the ability to recover fully-sampled image slice $I_{i,k}^u$ from under-sampled image slice $I_{i,k}^f$.

\subsection{Data Preprocess} \label{2.2}

The original dataset provided by the challenge organizer is composed of single-coil and multi-coil data. Each data is stored in .mat format, accompanied by striped masks of different acceleration factors (e.g., 4, 8, and 10). Under-sampled k-space data can be generated by covering the striped mask on the fully-sampled data. Based on the problem formulation, we first process fully-sampled and under-sampled k-space data to make data pairs (i.e., ground truth and input) for network training. Specifically, for single-coil data, we directly apply iFFT to obtain 2D image slices; for mult-coil data, we first apply RSS~\cite{rss} to aggregate k-space data from multiple coils and then apply iFFT to obtain 2D image slices.  
In other words, for a multi-coil input in shape $[t,s,c,h,w]$ (time-frame, slice, coil, height, width) or a single-coil input in shape $[t,s,h,w]$, the preprocessed data will be $t\times s$ 2D image slices in shape $[h,w]$.

We observe that blur noise is introduced by missing frequency information blocked by the striped mask. Hence, for padding these slices to a fixed shape, we choose to add zero padding to the k-space instead of the image space to keep the purify of the source of blur noise because padding zeros in the k-space will not bring new information from the frequency perspective while padding zeros in the image space will introduce unnecessary bias.

\begin{figure}[t]
\centering
\includegraphics[width=\textwidth]{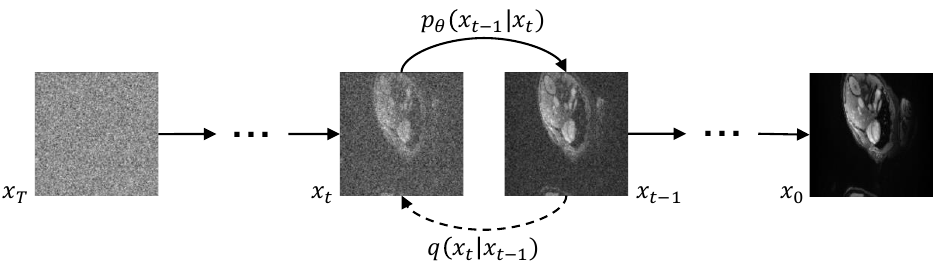}
\vspace{-7mm}
\caption{Forward and backward denoising diffusion processes.} 
\label{diff process}
\end{figure}

\subsection{Framework --- DiffCMR}
We briefly introduce the formulation of the diffusion model proposed in \cite{ddpm,ddim}. Diffusion models are generative models based on parametrized Markov chain and are comprised of forward and backward processes, as shown in Fig.~\ref{diff process}. The forward process iteratively transforms a clean image $x_0$ into a series of noisier images $\{x_1, x_2, ..., x_T\}$. The following formulation can express the iteration of the forward process:
\begin{equation} \label{eq1}
    q(x_t|x_{t-1}) = \mathcal{N}(x_t;\sqrt{1-\beta_t}x_{t-1}, \beta_tI_{n\times n}),
\end{equation}
where $\beta_t$ is a constant to define the ratio of adding Gaussian noise. The reverse process is parametrized by $\theta$ and can be simplified as:
\begin{equation}
    p_{\theta}(x_{t-1}|x_t) = \mathcal{N}(x_{t-1}; \epsilon_\theta(x_t, t), \sigma_t^2I_{n\times n}),
\end{equation}
where $\sigma_t$ is a fixed variance, and $\epsilon_\theta$ is a trained step estimation function.

Conditional generations with diffusion models \cite{cddpm,iddpm,segdiff} are formulated by letting backward process $p_\theta$ simultaneously manipulate the current step noisy image $x_t$ and the given condition, which allows the generation of images based on extra conditions without any additional learning. Here we merge the information of the under-sampled image and the current step denoising result by adding the extracted features from their corresponding encoders.

\begin{figure}[t]
\centering
\includegraphics[width=\textwidth]{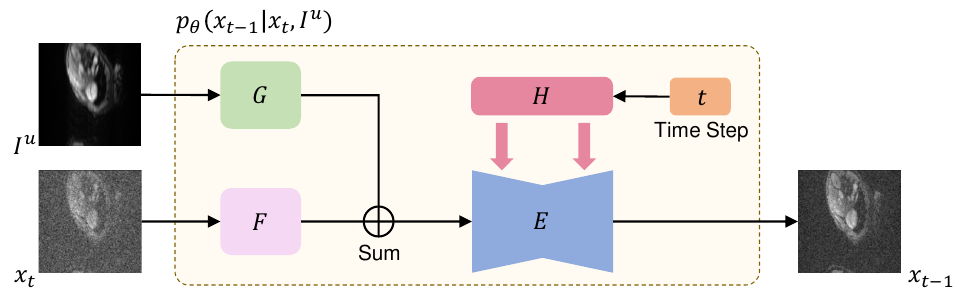}
\vspace{-7mm}
\caption{The figure above illustrates the pipeline of our proposed DiffCMR. $F$ and $G$ encode the noisy signal $x_t$ and the under-sampled image $I^u$, respectively. $H$ encodes timestamp $t$ to obtain timestamp embeddings. $E$ is a modified U-net that receives both summed features and timestamp embeddings for denoising.} 
\label{diffcmr}
\end{figure}

Our proposed DiffCMR, as illustrated in Fig.~\ref{diffcmr}, employs a conditional diffusion model that conditions its step estimation function $\epsilon_\theta$ on the aggregated information from both the input under-sampled image slice $I^u$ and the current step recovery $x_t$. In our architecture, the estimation function $\epsilon_\theta$ is a modified U-Net~\cite{unet} and can be further expressed as:
\begin{equation}
    \epsilon_\theta(x_t, I^u, t) = E(F(x_t) + G(I^u), H(t)),
\end{equation}
where $H$ encodes the current time step $t$ into timestamp embedding; $G$ and $F$ encode the under-sampled input slice $I^u$ and the current step denoising result $x_t$, respectively. $E$ is a modified U-Net encoder-decoder structure that receives summed features from $F$ and $G$, and estimates the noise for the current step. The current time step $t$ is embedded using a learned look-up table $H$ and inserted into layers of both the encoder and the decoder of network $E$ by summation.

\subsection{Training and Inference Procedure}
The training process for DiffCMR is demonstrated in Alg.~\ref{algtrain}. For each step, we sample a random data pair $(I^u, I^f)$ from the dataset $D_k$, a timestamp $t \in [1,T]$ from a Uniform distribution, and a noise $\epsilon$ from a standard Gaussian distribution. We then obtain the current step recovery $x_t$ by reparametrizing Eq.~\ref{eq1}:
\begin{equation}
    \alpha_t = 1 - \beta_t, \; \overline{\alpha}_t=\prod_{s=0}^t\alpha_s,  \; x_t = \sqrt{\overline{\alpha}_t}I^f + \sqrt{1-\overline{\alpha}_t}\epsilon. 
\end{equation}
Then according to our pipeline in Fig.~\ref{diffcmr}, $x_t$, $t$, and $I^u$ are sent through the networks $F$, $H$, $G$, and $E$ to obtain $\epsilon_\theta(x_t, I_i^u, t)$. Our training target is to minimize the term:
\begin{equation}
\Vert \epsilon - \epsilon_\theta(\sqrt{\overline{\alpha}_t}I^f + \sqrt{1-\overline{\alpha}_t}\epsilon, I_i^u, t)\Vert.
\end{equation}

Our inference process is described in Alg.~\ref{alginf}. As the procedure to recover $x_{t-1}$ includes an addition with a random sampled $z\sim \mathcal{N}(\mathbf{0},\mathbf{I_{n\times n}})$, which yields discrete random noise points in the denoising results. We adopt a multi-round ensembling inference strategy to diminish the noise points and stabilize the denoising results. The inference procedure on the same input is carried out for multiple rounds, and the final result is ensembled by taking the average. The effectiveness of this strategy is proved by the experiment results from the ablation study, see Tab.\ref{ablation} (b).

\begin{algorithm}[tb]
    \caption{Training Algorithm}
    \label{algtrain}
    \renewcommand{\algorithmicrequire}{\textbf{Input:}}
    \renewcommand{\algorithmicendfor}{}
    \begin{algorithmic}
        \REQUIRE total denoising steps $T$, under-sampled and fully-sampled image pair dataset $D_k=\{(I_i^u, I_i^f)\}_{i=1}^{N_k}$
        \REPEAT
        \STATE Sample $(I_i^u, I_i^f) \sim D_k, \; \epsilon \sim \mathcal{N}(\mathbf{0},\mathbf{I_{n\times n}}), \;t \sim \mathbf{Uniform(\{1,...,T\})}$
        \STATE $\beta_t = \frac{10^{-4}(T-t)+2\times10^{-2}(t-1)}{T-1}$, \; $\alpha_t = 1 - \beta_t$, \; $\overline{\alpha}_t = \prod_{s=0}^t\alpha_s$
        \STATE $x_t = \sqrt{\overline{\alpha}_t}I_i^f + \sqrt{1-\overline{\alpha}_t}\epsilon$
        \STATE Compute gradient $\nabla_\theta \Vert \epsilon - \epsilon_\theta(x_t, I_i^u, t)\Vert$
        \UNTIL convergence
    \end{algorithmic}
\end{algorithm}

\begin{algorithm}[tb]
    \caption{Inference Algorithm}
    \label{alginf}
    \renewcommand{\algorithmicrequire}{\textbf{Input:}}
    \renewcommand{\algorithmicendfor}{}
    \begin{algorithmic}
        \REQUIRE total denoising steps $T$, under-sampled image $I^u$, ensemble rounds $R$ 
        \FOR{$r=1,2,...,R$}
        \STATE Sample $x_{T,r} \sim \mathcal{N}(\mathbf{0},\mathbf{I_{n\times n}})$
        \FOR{$t=T,T-1,...,1$}
        \STATE Sample $z \sim N(\mathbf{0},\mathbf{I_{n\times n}})$
        \STATE $\beta_t = \frac{10^{-4}(T-t)+2\times10^{-2}(t-1)}{T-1}$
        \STATE $\alpha_t = 1 - \beta_t$, \; $\overline{\alpha}_t = \prod_{s=0}^t\alpha_s$, \; $\widetilde{\beta_t} = \frac{1-\overline{\alpha}_{t-1}}{1-\overline{\alpha}_t}\beta_t$
        \STATE $x_{t-1,r} = \alpha_t^{\frac{1}{2}}(x_t - \frac{1-\alpha_t}{\sqrt{1-\overline{\alpha}_t}}\epsilon_{\theta}(x_t,I^u,t)) + \mathbbm{1}_{[t>1]}\widetilde{\beta}_t^\frac{1}{2}z$
        \ENDFOR
        \ENDFOR
        \RETURN $\sum_{r=1}^Rx_{0,r}/R$
    \end{algorithmic}
\end{algorithm}

\section{Experiments}

\subsection{Dataset}

The CMRxRecon dataset is released in the MICCAI 2023 CMRxRecon Challenge, comprises cine reconstruction and T1/T2 mapping tasks. Both tasks have two coil types, each with 120 cases for training and 60 cases for validation. The training cases provide fully sampled k-space data and under-sampled k-space data with acceleration factors (AccFactor) 4, 8, and 10. The validation cases only provide under-sampled k-space data with acceleration factors 4, 8, and 10.

As described in Sec.~\ref{2.2}, we first preprocess the raw data by zero-padding the k-space to size 512$\times$512, and transforming it to 2D images with iFFT. Finally, we resize the image slices to 128$\times$128 to speed up the experiment process. Our local training-validation split is set by assigning images extracted from case P001 to P110 to the training set and those from case P111 to P120 to the validation set. We randomly shuffle the validation set and select the first 240 samples for inference. The detailed numbers of our preprocessed samples for training and validation are listed in Table~\ref{split}.

\begin{table}[t]
\centering
\setlength{\tabcolsep}{10pt}
\caption{Training and Validation pairs for both tasks and acceleration factors (AccFactor) 4, 8 and 10.}
\begin{tabular}{c|cc|cc|cc}
\toprule[1.2pt]
& \multicolumn{2}{c|}{AccFactor04} & \multicolumn{2}{c|}{AccFactor08} & \multicolumn{2}{c}{AccFactor10} \\ \cline{2-7}
& Train & Valid & Train & Valid & Train & Valid \\ \hline
Task 1 & 14304  & 1272  & 14304 & 1272  & 14304 & 1272 \\
Task 2 & 32904  & 2904  & 32904 & 2904  & 32904 & 2904 \\ 
\bottomrule[1.2pt]
\end{tabular}
\label{split}
\end{table}

\subsection{Implementation Details}
As for model architectures, we follow the conventions in \cite{segdiff} to build our model. The network $G$ has 10 Residual in Residual Dense Blocks~\cite{rrdb} and a depth of six. The number of channels was set to $[C,C,2C,2C,4C,4C]$ with $C=128$. The augmentation schemes include horizontal and vertical flips with a probability of 0.5. The training process took place with a batch size of 6 images at size 128 $\times$ 128. We used AdamW \cite{adamw} optimizer for all experiments. We used 100 diffusion steps for training and 1000 for inference. As there are two different tasks together with 3 acceleration factors, we trained the network 6 times to obtain 6 different sets of weight. The whole training and inference process is carried out on a single NVIDIA GeForce RTX 3090 GPU.

\begin{table}[t]
\centering
\caption{Experiment results for Task 1 - Cine Reconstruction}
\begin{tabular}{c|ccc|ccc|ccc}
\toprule[1.2pt]
\multicolumn{1}{l|}{}              & \multicolumn{3}{c|}{AccFactor04} & \multicolumn{3}{c|}{AccFactor08} & \multicolumn{3}{c}{AccFactor10} \\ \hline
\multicolumn{1}{l|}{} &
  \multicolumn{1}{c}{PSNR$\uparrow$} &
  \multicolumn{1}{c}{SSIM$\uparrow$} &
  \multicolumn{1}{c|}{NMSE$\downarrow$} &
  \multicolumn{1}{c}{PSNR$\uparrow$} &
  \multicolumn{1}{c}{SSIM$\uparrow$} &
  \multicolumn{1}{c|}{NMSE$\downarrow$} &
  \multicolumn{1}{c}{PSNR$\uparrow$} &
  \multicolumn{1}{c}{SSIM$\uparrow$} &
  \multicolumn{1}{c}{NMSE$\downarrow$} \\ \hline
RAW                                &    28.79       &     0.8150      &    0.2187      &    27.94       &      0.8082     &    0.2781      &     27.75      &    0.8113      &     0.2883     \\
U-Net~\cite{unet}                  &    33.11       &     0.9212      &    0.0673      &    33.31       &      \textbf{0.9460}     &    0.0646      &     32.71      &     \textbf{0.9391}      &     0.0740     \\
cGAN~\cite{cgan}                   &    33.85       &     \textbf{0.9435}      &    0.0573      &    33.14       &      0.9449     &    0.0669      &     32.56      &    0.9209      &     0.0760     \\
DiffCMR                            &    \textbf{36.10}       &     0.9277      &    \textbf{0.0346}      &    \textbf{34.85}       &      0.9061     &    \textbf{0.0457}      &     \textbf{34.47}      &    0.9016      &     \textbf{0.0493}     \\ \bottomrule[1.2pt]
\end{tabular}
\label{task1}
\end{table}

\begin{table}[t]
\centering
\caption{Experiment results for Task 2 - T1/T2 Mapping}
\begin{tabular}{c|ccc|ccc|ccc}
\toprule[1.2pt]
\multicolumn{1}{l|}{}              & \multicolumn{3}{c|}{AccFactor04} & \multicolumn{3}{c|}{AccFactor08} & \multicolumn{3}{c}{AccFactor10} \\ \hline
\multicolumn{1}{l|}{} &
  \multicolumn{1}{c}{PSNR$\uparrow$} &
  \multicolumn{1}{c}{SSIM$\uparrow$} &
  \multicolumn{1}{c|}{NMSE$\downarrow$} &
  \multicolumn{1}{c}{PSNR$\uparrow$} &
  \multicolumn{1}{c}{SSIM$\uparrow$} &
  \multicolumn{1}{c|}{NMSE$\downarrow$} &
  \multicolumn{1}{c}{PSNR$\uparrow$} &
  \multicolumn{1}{c}{SSIM$\uparrow$} &
  \multicolumn{1}{c}{NMSE$\downarrow$} \\ \hline
RAW                                &    28.17       &     0.8167      &    0.2003      &    27.15       &      0.8041     &    0.2578      &     27.17      &    0.8100      &     0.2711     \\
U-Net~\cite{unet}                  &    32.06       &     0.9340      &    0.0537      &    31.05       &      \textbf{0.9286}     &    0.0695      &     29.49      &     0.9106     &     0.0987     \\
cGAN~\cite{cgan}                   &    32.67       &     \textbf{0.9460}      &    0.0488      &    30.65       &      0.8899     &    0.0805      &     31.38      &    \textbf{0.9304}      &    0.0655      \\
DiffCMR                            &    \textbf{34.60}       &     0.9071      &    \textbf{0.0372}      &    \textbf{33.17}       &      0.8937     &    \textbf{0.0537}      &     \textbf{33.04}      &    0.8941      &     \textbf{0.0536}     \\ \bottomrule[1.2pt]
\end{tabular}
\label{task2}
\end{table}

\subsection{Results}
As the online validation platform limits the daily attempts to 3 trials per task, we perform the validation and report the results on our local split dataset to speed up the upgrading procedure of our method. For the fairness of comparison, all the experiments are trained and validated with the same split, input resolution, and the same data augmentation scheme. The evaluation metrics for our experiments are peak signal-to-noise ratio (PSNR), structural similarity index measure (SSIM), and normalized mean square error (NMSE).

\begin{figure}[t] 
\centering
\includegraphics[width=\textwidth]{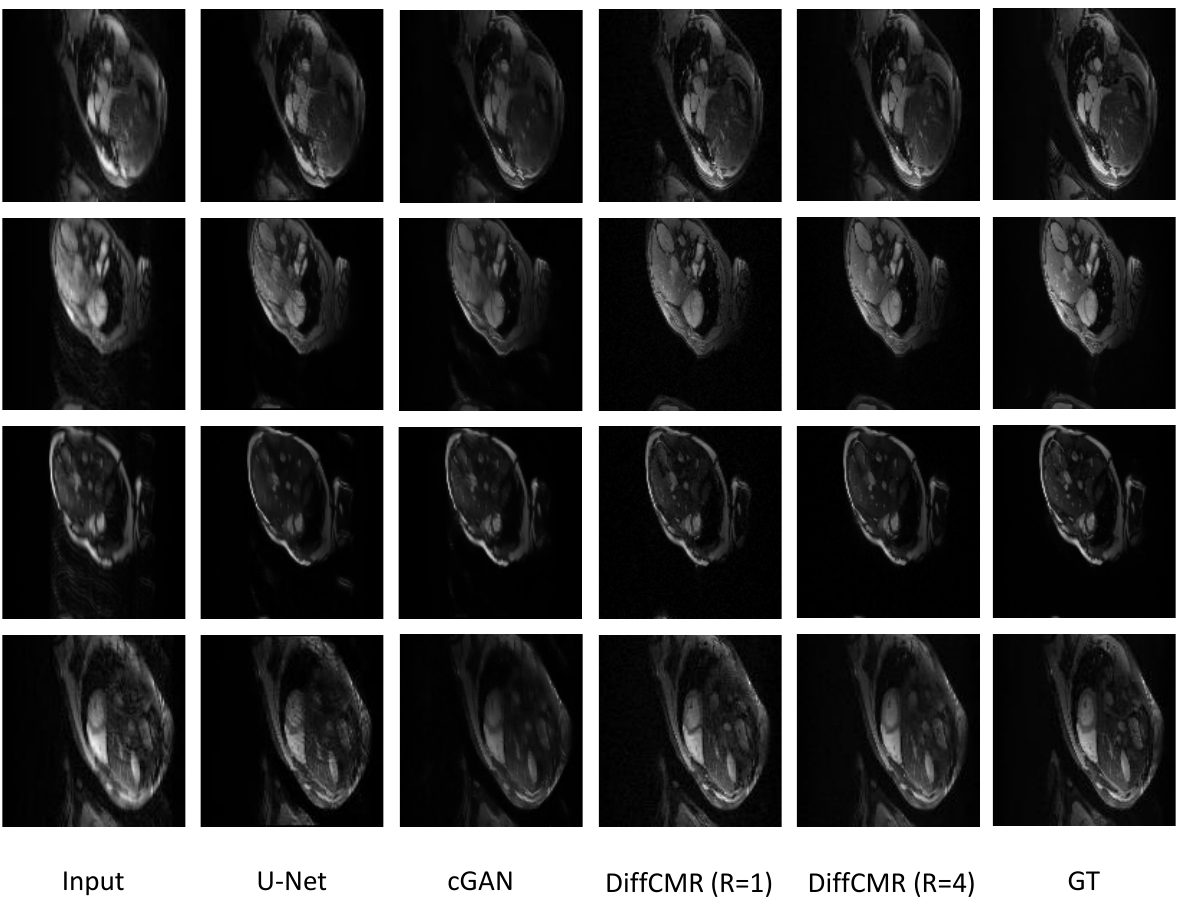}
\vspace{-5mm}
\caption{Qualitative comparisons between raw input, results of U-Net~\cite{unet}, results of cGAN~\cite{cgan}, results of DiffCMR with single-round ensembling inference ($R=1$), results of DiffCMR with multi-round ensembling inference ($R=4$), and ground truth.} 
\label{visual}
\end{figure}

We compare our proposed DiffCMR with U-Net \cite{unet} and cGAN \cite{cgan}. Qualitative visualization comparisons are shown in Fig.~\ref{visual}. Quantitative results for Task1 and Task2 are listed in Tab.~\ref{task1} and Tab.~\ref{task2}, where RAW means the direct comparison results between the input under-sampled image slices and the fully-sampled image slices. As can be seen, our method outperforms both baseline methods across most tasks and acceleration factors.

\section{Ablation Study}



We evaluate two alternatives of hyper-parameters in our DiffCMR method at the inference stage. The first variant determines the number of inference diffusion steps. The second variant determines the number of inference ensembling rounds. These variant experiments were carried out on the data from Task 1 with an acceleration factor equal to 4.

\begin{table}[t]
\caption{Ablation study on Task 1-AccFactor04. (a) Results with different inference diffusion steps. (b) Results with different inference ensembling rounds.}
\vspace{4pt}
\centering
\begin{minipage}[t]{0.48\textwidth}
\centering
\setlength{\tabcolsep}{4.4pt}
\begin{tabular}{c|ccc}
\toprule[1.2pt]
\#Steps $T$ & PSNR$\uparrow$  & SSIM$\uparrow$   & NMSE$\downarrow$   \\ \hline
Raw    & 28.79 & 0.8150 & 0.2187 \\
20     & 25.52 & 0.3664 & 0.3830 \\
100    & 29.86 & 0.6006 & 0.1393 \\
500    & 35.39 & 0.9010 & 0.0399 \\
1000   & \textbf{36.10} & \textbf{0.9277} & \textbf{0.0346} \\ \bottomrule[1.2pt]
\end{tabular}
\vspace{-8.5pt}
\caption*{(a)}

\end{minipage}
\begin{minipage}[t]{0.48\textwidth}
\centering
\setlength{\tabcolsep}{4.4pt}
\begin{tabular}{c|ccc}
\toprule[1.2pt]
\#Rounds $R$ & PSNR$\uparrow$  & SSIM$\uparrow$   & NMSE$\downarrow$   \\ \hline
Raw     & 28.79 & 0.8150 & 0.2187 \\
1       & 33.89 & 0.8491 & 0.0561 \\
2       & 35.25 & 0.8994 & 0.0417 \\
4       & 36.10 & 0.9277 & 0.0346 \\
8       & \textbf{36.68} & \textbf{0.9430} & \textbf{0.0305} \\ \bottomrule[1.2pt]
\end{tabular}
\vspace{-12pt}
\caption*{(b)}
\end{minipage}
\label{ablation}
\end{table}

\vspace{6pt}
\noindent
\textbf{Varying the number of inference diffusion steps $T$.} \quad In this part, we set the ensembling rounds $R=4$ and explore the effect of inference diffusion steps on the denoising quality. Quantitative results are shown in Tab.~\ref{ablation}(a). As can be observed, the denoising performance is positively correlated with $T$ and starts to outperform the raw input at around $T=100$. We choose $T=1000$ in all other experiments.

\vspace{6pt}
\noindent
\textbf{Varying the number of inference ensembling rounds $R$.} \quad In this part, we set the diffusion steps $T=1000$ and explore the effect of inference ensembling rounds on the denoising quality. Quantitative results are shown in Tab.~\ref{ablation}(b), and Qualitative comparisons between $R=1$ and $R=4$ are visualized in Fig.~\ref{visual}. The results show a positive correlation between denoising effectiveness and $R$ with a diminishing marginal effect when $R$ is large. Therefore, we set $R=4$ in all other experiments for a reasonable performance to runtime tradeoff.

\section{Conclusion}
In this paper, we present DiffCMR, a conditional DDPM-based approach for high-quality MRI reconstruction from under-sampled k-space data.
Our framework receives conditioning signals from the under-sampled MRI image slice at each denoising diffusion step and generates the corresponding fully-sampled MRI image slice. 
In addition, we adopt the multi-round ensembling strategy during inference which largely enhances the stableness of our approach.
Experiment results show our DiffCMR method outperforms the existing popular denoisers qualitatively and quantitatively.
In conclusion, our proposed DiffCMR offers a novel perspective for handling fast MRI reconstruction problems and demonstrates impressive robustness.

\vspace{6pt}
\noindent
\textbf{Acknowledgement.} 
This work was supported by the Hong Kong Innovation and Technology Fund under Projects PRP/041/22FX.

\bibliographystyle{splncs04}
\bibliography{cmr.bib}

\end{document}